\input epsf
\documentstyle{mn}
\newif\ifAMStwofonts

\def\lesssim{\mathrel{\hbox{\rlap{\hbox{\lower4pt\hbox{$\sim$}}}\hbox{$<$}}}}
\def\gtrsim{\mathrel{\hbox{\rlap{\hbox{\lower4pt\hbox{$\sim$}}}\hbox{$>$}}}}
\def\apj{ApJ}

\def\aap{A\&\hskip-1pt A}

\def\mnras{MNRAS}

\def\araa{ARA\&\hskip-1pt A}
\title[Caustic-Crossing Intervals]
      {Distribution of Caustic-Crossing Intervals for \\
       Galactic Binary-Lens Microlensing Events}

\author[Han, C., Park, S.-H., \& Lee, Y.-S.]
       {Cheongho Han, Seong-Hong Park, \& Yong-Sam Lee\\
        Department of Astronomy \& Space Science, \\
        Chungbuk National University, Chongju, Korea 361-763\\
        cheongho,parksh,leeys@astronomy.chungbuk.ac.kr}

\date{Accepted
      Received }

\pagerange{\pageref{firstpage}--\pageref{lastpage}}

\begin{document}

\maketitle

\label{firstpage}

\begin{abstract}
Detection of caustic crossings of binary-lens gravitational microlensing events 
is important because by detecting them one can obtain useful information both 
about the lens and source star.  In this paper, we compute the distribution of 
the intervals between two successive caustic crossings, $f(t_{\rm cc})$, for 
Galactic bulge binary-lens events to investigate the observational strategy 
for the optimal detection and resolution of caustic crossings.  From this 
computation, we find that the distribution is highly skewed toward short 
$t_{\rm cc}$ and peaks at $t_{\rm cc}\sim 1.5$ days.  For the maximal detection 
of caustic crossings, therefore, prompt initiation of followup observations for 
intensive monitoring of events will be important.  We estimate that under the 
strategy of the current followup observations with a second caustic-crossing 
preparation time of $\sim 2$ days, the fraction of events with resolvable 
caustic crossing is $\sim 80\%$.  We find that if the followup observations 
can be initiated within 1 day after the first caustic crossing by adopting 
more aggressive observational strategies, the detection rate can be improved 
into $\sim 90\%$.
\end{abstract}


\section{Introduction}

Several groups are searching for massive astronomical compact objects (MACHOs) 
by monitoring light variations of stars caused by gravitational microlensing
towards the Galactic bulge and the Magellanic Cloud fields (EROS: Aubourg et 
al.\ 1993; MACHO: Alcock et al.\ 1993; OGLE: Udalski et al.\ 1993; DUO: Alard 
\& Guibert 1997). With their efforts, more than 400 candidate events have been 
detected to date.

Among these events, a considerable number of events are believed to be caused 
by binary lenses (Udalski et al.\ 1994a; Dominik \& Hirshfeld 1994, 1996;
Mao \& Di Stefano 1995; Alard, Mao, \& Guibert 1995; Bennett et al.\ 1996; 
Rhie \& Bennett 1996; Alcock et al.\ 1999b; Afonso et al.\ 2000).
The light curve of a binary-lens event differs from 
that of a single-lens event. The most distinctive feature of a binary-lens 
event light curve occurs when a source star crosses a lens caustic. When the 
source star is located on the caustic, the amplification becomes very large. 
Therefore, the light curve of a caustic-crossing binary-lens event is 
characterized by a sharp spike.  On the other hand, if a binary-lens event 
does not involve a caustic crossing, the resulting light curve exhibits a 
relatively small deviation from that of a single-lens event (Mao \& 
Paczy\'nski 1991; Dominik 1998).

Detection of caustic-crossing binary-lens events is important because they 
can provide useful information both about the lens and source star. First, a 
caustic-crossing event provides an opportunity to measure how long it takes for
the caustic line to transit the face of the source star. By using the source 
crossing time along with an independent determination of the source star size, 
one can determine the lens proper motion with respect to the source star
(Afonso et al.\ 1998; Albrow et al.\ 1999a; Alcock et al.\ 1999a).
With the determined lens proper motion, one can significantly better constrain 
the physical parameters of the lens (Gould 1994; Nemiroff \& Wickramasinghe 
1994; Witt \& Mao 1994; Peng 1997).
Second, a caustic-crossing event can also be used to determine the surface 
structure of the source star such as the radial brightness profiles (Gaudi 
\& Gould 1999; Albrow et al.\ 1999b, 2000; Afonso et al.\ 2000) and spots 
(Han et al.\ 1999). 
To obtain these useful information from the light curve of a binary-lens event, 
the event should be monitored with high time resolution. However, under the  
nightly monitoring observational strategy of the current lensing experiments, 
it is difficult to construct light curves with resolution high enough to  
resolve caustics.

However, the caustic crossing can be resolved with the help of alert system 
based on real time observations (Alcock et al.\ 1996, 1997; Udalski et al.\ 
1994b) and subsequent high time resolution followup observations (GMAN:  
Pratt et al.\ 1996; PLANET: Albrow et al.\ 1998; MPS: Rhie et al.\ 1998). 
Since the caustics of a binary-lens event form a closed curve, the source  
star of a caustic-crossing event crosses the caustic line at least twice. 
Although the first caustic crossing is unlikely to be resolved due to its 
short duration, it can be inferred from the enhanced amplification. 
Then, if followup observations can be prepared before the second caustic 
crossing, dense enough sampling through the second caustic will be possible. 
Therefore, it is important to estimate the distribution of the intervals 
between caustic crossings (caustic-crossing intervals, $t_{\rm cc}$) for the 
optimal detection and resolution of caustic crossings.

The distribution of caustic-crossing intervals, $f(t_{\rm cc})$, was 
computed by Honma (1999) for events expected towards the 
Magellanic Cloud field.  His intention for the computation of $f(t_{\rm cc})$ 
was to demonstrate the detection bias against events with short $t_{\rm cc}$,
for which the second caustic crossings are more likely to be missed. For this 
purpose, it was enough to determine $f(t_{\rm cc})$ based on the typical 
values of the lens mass and the Einstein ring radius crossing time (Einstein 
time scale, $t_{\rm E}$). However, for the investigation of the optimal 
observational strategy to detect and resolve caustic crossings, it is required 
to determine $f(t_{\rm cc})$ based on the distribution of Einstein time scales 
expected from detailed models of the physical and dynamical distributions of 
lens matter and binary mass function. In addition, since majority of 
caustic-crossing binary-lens events have been detected towards the Galactic 
bulge (Alcock 1999b), constructing $f(t_{\rm cc})$ for 
these events is desirable.

In this paper, we compute the distribution of the intervals between two 
successive caustic crossings, $f(t_{\rm cc})$, for Galactic bulge binary-lens 
events to investigate the observational strategy for the optimal detection 
and resolution of caustic crossings.  From this computation, we find that the 
distribution is highly skewed toward short $t_{\rm cc}$ and peaks at 
$t_{\rm cc}\sim 1.5$ days.  For the maximal detection of caustic crossings, 
therefore, prompt initiation of followup observations for intensive monitoring 
of events will be important.  We estimate that under the strategy of the 
current followup observations with a second caustic-crossing preparation time 
of $\sim 2$ days, the fraction of events with resolvable caustic crossing is 
$\sim 80\%$.  We find that if the followup observations can be initiated 
within 1 day after the first caustic crossing by adopting more aggressive 
observational strategies, the detection rate can be improved into $\sim 90\%$.

\begin{figure}
\epsfysize=10cm
\vskip-1cm
\centerline{\epsfbox{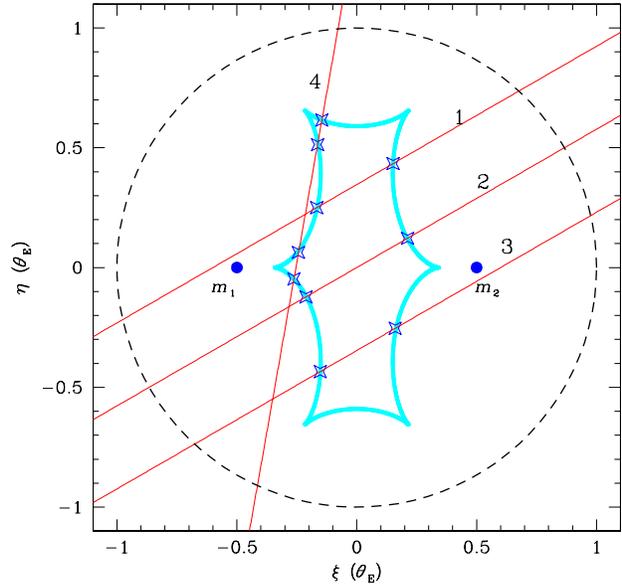}}
\vskip-1cm
\caption{
Caustic crossings for various source star trajectories (thin solid straight 
lines) for an example event caused by a binary system with a separation 
$\ell=1.0$ and a mass ratio $q=1.0$. The caustics are represented by a thick 
sold line. The location of the lenses ($m_1$ and $m_2$) are marked by dots 
and the circle drawn by a dashed line represents the combined Einstein ring.
}
\end{figure}

\section{Caustic-Crossing Intervals}

\subsection{Binary-lens Events}

When the lengths are normalized to the combined Einstein ring radius, the 
lens equation in complex notations for a binary-lens system is represented by 
$$
\zeta = z + {m_{1} \over \bar{z}_{1}-\bar{z}} 
+ {m_{2} \over \bar{z}_{2}-\bar{z}},
\eqno(1)
$$
where $m_1$ and $m_2$ are the mass fractions of individual lenses (and thus 
$m_1+m_2=1$), $z_1$ and $z_2$ are the positions of the lenses, $\zeta = \xi +
i\eta$ and $z=x+iy$ are the positions of the source and images, and $\bar{z}$ 
denotes the complex conjugate of $z$ (Witt 1990). The combined 
Einstein ring radius is related to the total mass $M$ of the binary system 
and its location between the observer and the source star by
$$
r_{\rm E} =\left( {4GM\over c^2} {D_{ol}D_{ls}\over D_{os}}\right)^{1/2}, 
\eqno(2)
$$
where $D_{ol}$, $D_{ls}$, and $D_{os}$ are the separations between the 
observer, lens, and source star. Since the individual microlensing images 
cannot be resolved due to the small separations between them, the amplification 
of the binary-lens event is given by the sum of the amplifications of the 
individual images, $A_i$, which are given by the Jacobian of the transformation
(1) evaluated at the image position, i.e.\  
$$
A_i = \left({1\over \vert {\rm det}\ J\vert} \right)_{z=z_i};
\qquad {\rm det}\ J = 1-{\partial\zeta\over\partial\bar{z}}
{\overline{\partial\zeta}\over\partial\bar{z}}.
\eqno(3)
$$
The caustic refers to the source position on which the amplification becomes
infinity, i.e.\ ${\rm det}\ J=0$.\footnote{The actually observed amplification 
is finite because the source star is not a point source. The amplification for 
an extended source is the weighted mean of the amplification factors over the 
surface of the source, leading to a finite value (Schneider \& Weiss 
1986).} The set of caustics form a closed curve, called 
cautsics. The caustics take various shape and size depending on the binary 
mass ratio $q=m_1/m_2$ and the projected binary separation $\ell$ normalized 
by the angular Einstein ring radius $\theta_{\rm E}=r_{\rm E}/D_{ol}$. In 
Figure 1, we present the caustics (a thick solid curve) of an example 
binary-lens event with $\ell=1.0$ and $q=1.0$. In the figure, the locations of 
the lenses ($m_1$ and $m_2$) are marked by dots and the circle with its center 
at the center of mass of the binary system drawn by a dashed line represents 
the combined Einstein ring. For more example caustics of binary-lens events 
with various values of $\ell$ and $q$, see Figure 1 of Han 
(1999).

\subsection{Distribution of Caustic-Crossing Separations}

The caustic-crossing interval is proportional to the separation between two 
successive caustic-crossing points (caustic-crossing separation, 
$\theta_{\rm cc}$). Due to the dependency of the shape and size of the caustics 
on the binary separation and the mass ratio, the caustic-crossing separation 
and the resulting caustic-crossing interval depend on the values of $\ell$ 
and $q$. In addition, since $t_{\rm cc}$ is scaled by the Einstein time scale, 
which depends on the physical lens parameters of the mass, the location, and 
the lens-source transverse speed $v$ by
$$
t_{\rm E}={r_{\rm E}\over v} 
= \left( {4GM\over c^2v^2} {D_{ol}D_{ls}\over D_{os}}\right)^{1/2},
\eqno(4)
$$
the caustic-crossing interval depends also on these parameters. Therefore, 
for the determination of $f(t_{\rm cc})$, one should consider binary-lens 
events which are expected for all combinations of $\ell$ and $q$ over the 
entire range of the physical lens parameters.

To determine $f(t_{\rm cc})$, we first determine the distribution of the 
caustic-crossing separations $f(\theta_{\rm cc}; \ell, q)$, which is 
expected for events caused by a binary lens with $\ell$ and $q$. By 
defining a binary-lens event as a close lens-source encounter within 
the combined Einstein ring\footnote{Some binaries form their caustics 
outside their combined Einstein ring. However, the size of the outer 
caustics is usually very small compared with the combined Einstein ring, 
implying that the probability of the outer caustic crossing will be very 
small. Therefore, determination of $f(\theta_{\rm cc}; \ell, q)$ will not 
be seriously affected by our definition of a binary-lens event.}, we produce 
a large number of source trajectories that pass through the combined 
Einstein ring. The source star trajectory orientations with respect to the 
projected binary axis, $\psi$, and the impact parameter, $\beta$, are 
randomly selected in the ranges of $0\leq\psi\leq 2\pi$ and $0\leq\beta\leq 1$,
respectively. If the source star trajectory crosses the lens caustics, we 
then find the crossing points and measure the separation $\theta_{\rm cc}$ 
in units of $\theta_{\rm E}$. In most cases, the source crosses the caustics 
twice, e.g.\ the three trajectories numbered by 1, 2, and 3 in Figure 1. 
But due to the concavity of the caustics, crossings can occur more than 
twice, e.g.\ the trajectory numbered by 4 in Figure 1. For these cases, 
we measure $\theta_{\rm cc}$ between every pair of two successive crossing 
points.

\begin{figure}
\epsfysize=9.0cm
\vskip-1cm
\centerline{\epsfbox{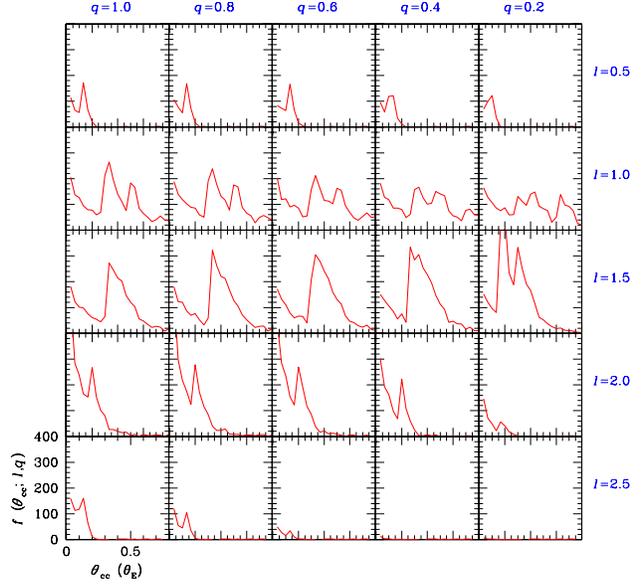}}
\caption{
The distributions of the normalized caustic-crossing separations for events 
caused by binary systems with various separations $\ell$ (also normalized by 
$\theta_{\rm E}$) and mass ratios $q$. The individual distributions are 
arbitrarily normalized, but they are relatively scaled so that the area under 
each distribution is proportional to the caustic-crossing probability.
}
\end{figure}

In Figure 2, we present the distributions $f(\theta_{\rm cc};\ell,q)$ for 
events caused by binaries with various projected separations and mass ratios. 
The individual distributions are arbitrarily normalized, but they are 
relatively scaled so that the area under each distribution is proportional 
to the caustic-crossing probability.  From the figure, one finds that the 
distribution $f(\theta_{\rm cc};\ell,q)$ depends strongly, both in the scale 
(i.e.\ the caustic-crossing probability) and the shape, on the binary 
separation. In the scale, the caustic probability becomes maximum for 
binary-lens events with $\ell\sim 1.5$. This is because caustics at around 
this binary separation form the largest curve.  With the same reason, the 
distribution peaks at relatively large caustic-crossing separations, e.g.\ 
at around $\theta_{\rm cc}=0.4$ for binary-lens events with $\ell\sim 1.5$. 
As the separation becomes smaller or larger, the caustics becomes smaller, 
making the caustic-crossing probability smaller and the distribution peaks 
at small value of $\theta_{\rm cc}$. On the other hand, the dependency of 
$f(\theta_{\rm cc};\ell,q)$ on the binary mass ratio both in the scale and 
the shape is relatively weak (Gaudi \& Gould 1999).

With the obtained distribution of $f(\theta_{\rm cc};\ell,q)$, we then obtain 
the combined distribution of caustic-crossing separations by
$$
f(\theta_{\rm cc}) = \int_0^1 dq f(q)
\int_0^\infty db f(\theta_{\rm cc};\ell,q) f(b)
\delta (\theta_{\rm cc}-\theta_{\rm cc}'),
\eqno(5)
$$
where $f(q)$ and $f(b)$ respectively are the distributions of the binary mass 
ratios and the intrinsic binary separations $b$ in units of $\theta_{\rm E}$, 
and $\delta$ denote the delta function. The projected binary separation $\ell$ 
is obtained from the intrinsic separation $b$ by assuming that the intrinsic 
binary separation vector ${\bf b}$ is randomly oriented onto the sky. The 
notation $\theta_{\rm cc}'$ represents the caustic-crossing separation 
obtained under the assumption of random orientation of ${\bf b}$. For the 
distribution $f(q)$, we adopt the distribution determined by Trimble 
(1990).

\subsection{Distribution of Caustic-Crossing Intervals}

Once the distribution of caustic-crossing separations $f(\theta_{\rm cc})$ is 
obtained, the distribution of the caustic-crossing intervals expected for 
events caused by binaries with a {\it constant} mass $M$ is obtained by
$$
f(t'_{\rm cc})=\int_{0}^{d_{\rm max}} dD_{os} \rho(D_{os}) \int_{0}^{D_{os}} 
dD_{ol} \rho(D_{ol})\left( {D_{ol}D_{ls}\over D_{os}} \right)^{1/2}
$$
$$
\times 
\int dv_{y}\int dv_{z} v f(v_{y},v_{z})
\int_{0}^{\infty} d\theta_{\rm cc} f(\theta_{\rm cc}) \delta 
\left( t'_{\rm cc} - \theta_{\rm cc} t_{\rm E} \right),
\eqno(6)
$$
where $\rho(D_{os})$ and $\rho(D_{ol})$ are the mass density of source stars 
and lenses along the line of sight toward the Galactic bulge, $d_{\rm max}$ 
is the maximum extent of the source star distribution, $(v_y,v_z)$ are the 
two components, parallel and normal to the Galactic plane, of the 
lens-source transverse velocity {\bf v}, and $f(v_y,v_z)$ represents their 
distribution.  In the equation, the factors $(D_{ol}D_{ls}/D_{os})^{1/2}$ and 
$v$ are included to weight $f(t'_{\rm cc})$ by the cross section of the 
lens-source encounter, i.e.\ the Einstein ring radius 
$r_{\rm E}\propto (D_{ol}D_{ls}/D_{os})^{1/2}$, and the transverse speed.

For the matter distributions of the Galactic bulge, we adopt a `revised 
COBE' model, which is based on a triaxial COBE model (Dwek et al.\ 
1995) except the central part of the bulge. For the inner 
$\sim 600\ {\rm pc}$ of the bulges, we adopt a high central density Kent
(1992) model which better matches with observations in this 
region. For the matter distributions of the Galactic disk, we adopt a double 
exponential disk model with vertical and radial scale heights of $h_R=325\ 
{\rm pc}$ and $h_z=3.5\ {\rm kpc}$ (Bahcall 1986).
The transverse velocity distributions for both Galactic bulge and disk lenses 
are modelled by a Gaussian of the form
$$
f(v_i) \propto \exp \left[-  
{\left(v_i - \bar v_i \right)^{2} \over {2\sigma_i}^2} 
\right];\qquad  i \in y,z,
\eqno(7)
$$
where the means and the standard deviations for individual transverse velocity 
components are $({\bar v_y},\sigma_y)$=(220,30) ${\rm km\ s}^{-1}$ and 
$({\bar v_z},\sigma_z)$=(0,20) ${\rm km\ s}^{-1}$ for the disk lenses, and
$({\bar v_y},\sigma_y)$=(220,93) ${\rm km\ s}^{-1}$ and  
$({\bar v_z},\sigma_z)$=(0,79) ${\rm km\ s}^{-1}$ for the bulge lenses, 
respectively. For the detailed discussion both about the matter and the 
transverse velocity distributions, see Han \& Gould (1996).

The expected distributions of $f(t'_{\rm cc})$ for events caused by various 
values of the binary mass are presented in the upper panel of Figure 3.
In the lower panel, we also present the integrated probability for events 
with $t > t'_{\rm cc}$, i.e.\ $P(t>t'_{\rm cc})=1-\int_0^t f(t'_{\rm cc}) 
dt'_{\rm cc}/ \int_0^\infty f(t'_{\rm cc}) dt'_{\rm cc}$. In each panel, to 
better show the distribution in the short caustic-crossing interval region, 
we expand the region and present in a separate small box. From the figure, 
one finds that as the binary-lens mass decreases, the distribution becomes 
narrower and peaks at shorter $t_{\rm cc}$.

\begin{figure}
\epsfysize=12.0cm
\vskip-0.5cm
\centerline{\epsfbox{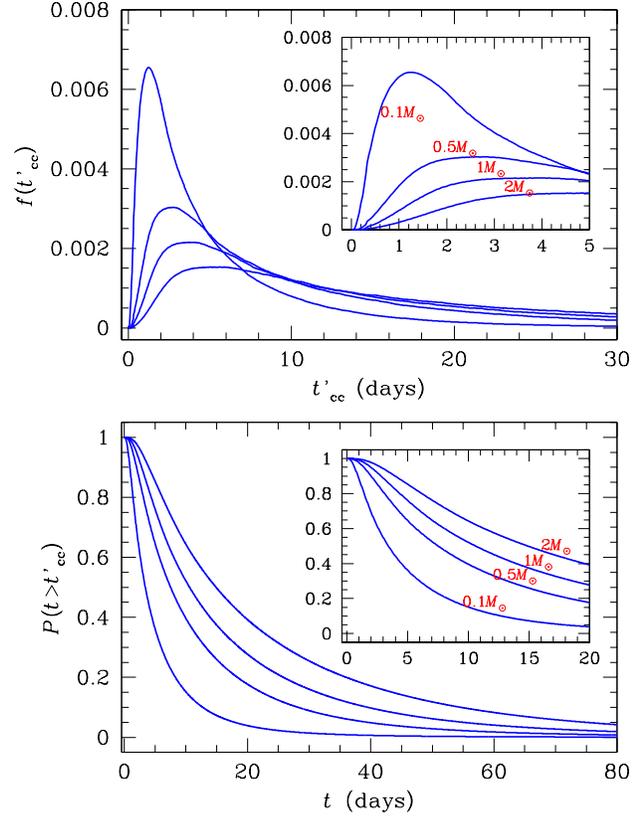}}
\vskip-0.5cm
\caption{
Upper panel:
The distributions of caustic-crossing intervals for Galactic bulge events 
caused by binaries with various constant masses.
Lower panel: 
The integrated probability for caustic-crossing events with $t>t'_{\rm cc}$. 
In each panel, to better show the short $t'_{\rm cc}$ region, we expand the 
region and presented in a separate small box.
}
\end{figure}

Binary-lens events are caused by Galactic binaries with various masses. Then
the final distribution of the caustic-crossing intervals is obtained by 
convolving $f(t'_{\rm cc})$ with the mass function of binary lenses
$\Phi_{\rm bi} (M)$, i.e.
$$
f(t_{\rm cc}) = \int dM M^{1/2} \Phi_{\rm bi}(M) \int_{0}^{\infty}
dt'_{\rm cc} f(t'_{\rm cc}) \delta (t_{\rm cc} - M^{1/2} t'_{\rm cc}).
\eqno(8)
$$
The factor $M^{1/2}$ in the first integrand is included to weight 
$f(t_{\rm cc})$ by the cross-section of lens-source encounter 
($r_{\rm E}\propto M^{1/2}$), while the same factor in the second integrand 
is included because the caustic crossing-crossing interval is scaled by the 
Einstein time scale ($t_{\rm E}\propto M^{1/2}$).

\begin{figure}
\epsfysize=9.5cm
\vskip-0.3cm
\centerline{\epsfbox{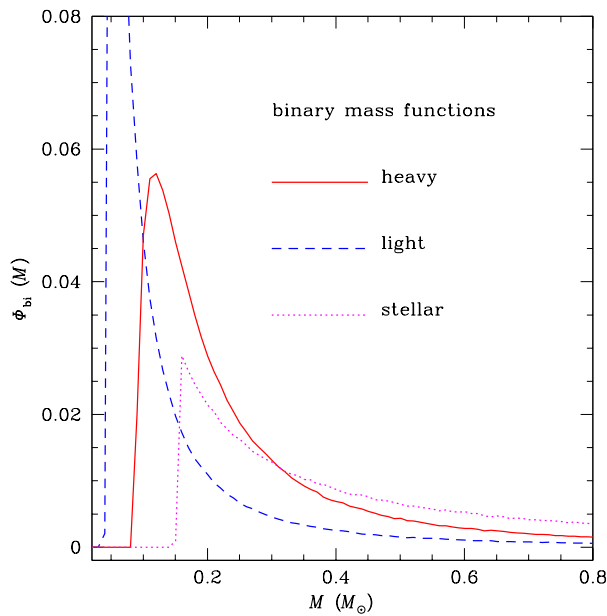}}
\vskip-0.2cm
\caption{
The model binary-lens mass functions that are used for the determination of 
the caustic-crossing interval distributions.
}
\end{figure}

The mass function of binary lenses is very uncertain due to the observational 
difficulties in identifying binaries by resolving their components combined 
with our ignorance about the mechanism of binary formation. Furthermore, since 
binary-lens microlensing events can be caused by binaries which are composed 
of dark components (either one or both), construction of $\Phi_{\rm bi}(M)$  
becomes even more complicated. Though many possibilities have been proposed, 
the scenarios for the formation of binaries can be classified into two 
categories. The first one is that binaries are formed during the star formation
stage through  fragmentation of a collapsing object (Norman \& Wilson 1970; 
Boss 1988). In the other scenario, by contrast, both 
binary components form independently and then later combine through some sort 
of capture mechanism (Fabian, Pringle, \& Rees 1975; Press \& Teukolsky 1997; 
Ray, Kembhavi, \& Anita 1987). Due to the 
difference in the binary formation mechanism, the mass functions expected 
from the two scanarios would be different each other. Under the formation 
environments of the former scenario, the most probable binary-lens mass 
function would be similar to that of single lenses. On ther other hand, under 
the circumstances of the latter binary formation scenario, the two lens masses 
would be drawn independently from the same mass function, which is similar to 
the single-lens mass function. For the construction of $f(t_{\rm cc})$, 
therefore, we test both binary mass function models expected under the two 
cases of binary formation scenario. The first binary mass function is modeled 
by a power law with a mass cutoff, i.e.\ 
$$
\Phi_{\rm bi}(M)\propto M^{-p} \Theta (m-m_{\rm cut}),
\eqno(9)
$$
where $\Theta$ is a heavy side step function. For the values of the power 
and the mass cutoff, we adopt $p=2.1$ and $m_{\rm cut}=0.04\ M_\odot$, 
following the best-fitting values determined from the distribution of 
Einstein time scales of 51 Galactic bulge single-lens events by Han \& 
Gould (1996). The second model of binary mass function is 
constructed by selecting masses of individual binary components from the 
same mass function in equation (9) and then combining the two masses. 
Since two masses are combined to yield a single binary mass, binaries 
following the latter mass function tend to be heavier than the binary lens 
masses following the former mass function.  We distinguish the two models 
by calling them `heavy' and `light' binary mass functions.  Note that since 
the mass cutoff of both the heavy and light binary mass functions is less 
than the hydrogen-burning limit of $0.08\ M_\odot$, dark component of lenses 
are included in these mass functions.  In the third model, we test binary 
mass function which is constructed under the assumption that lenses are 
composed of only stars: `stellar' mass function.  We construct the stellar 
mass function by adopting the mass function determined by Zoccali et al.\ 
(2000) from observations of Galactic bulge stars by 
using the Hubble Space Telescope (HST) plus the Near Infrared Camera and 
Multi Object Spectrometer (NICOMOS).  The stellar binary-lens mass function 
extends down to $0.15\ M_\odot$.  In Figure 4, we present the model 
binary-lens mass functions. 
 
\begin{figure}
\epsfysize=12.0cm
\centerline{\epsfbox{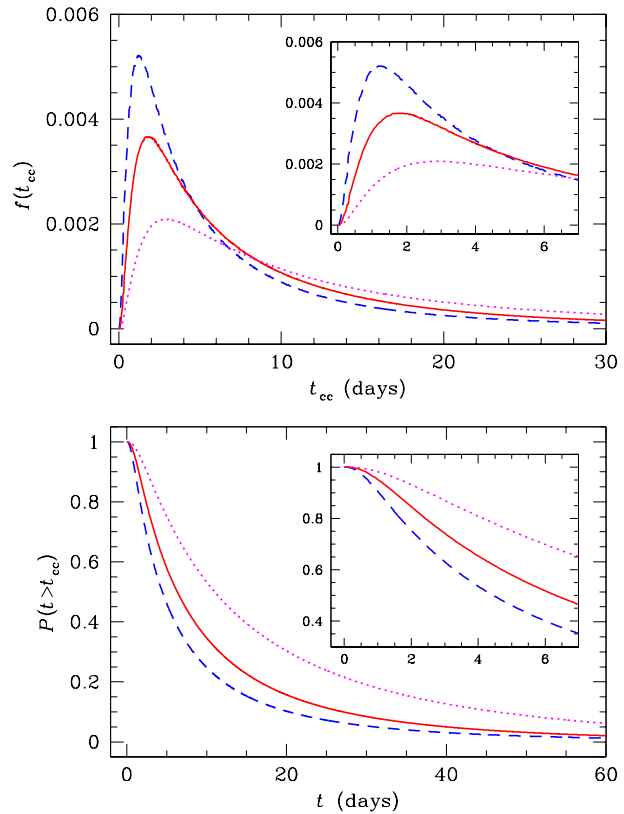}}
\vskip-0.5cm
\caption{
The resulting distributions of caustic-crossing intervals (lower left panel)
and the integrated probabilities with $t > t_{\rm cc}$ for Galactic bulge 
binary-lens events determined for the two model binary-lens mass functions 
in the upper panel.
}
\end{figure}

In Figure 5, we present the finally determined distribution of 
caustic-crossing intervals and integrated probability with $t> t_{\rm cc}$. 
From the figure, one finds that the distributions $f(t_{\rm cc})$ are 
highly skewed toward short $t_{\rm cc}$ regardless of the assumed mass 
functions.  The peak of the distribution occurs at $\sim 1.5\ M_\odot$ 
although ther exist slight variations depending on the assumed binary-lens 
mass functions.  As a result, majority (70\%--90\%) of binary-lens events 
expected to be detected towards the Galactic bulge will have caustic-crossing 
intervals shorter than $t_{\rm cc}=20$ days.

\section{Discussion}

In the current strategy, intensive followup observations of the second 
caustic crossing of a binary-lens event can be initiated within $\sim 2$ 
days after the first crossing.  The primary search teams alert on the first 
caustic crossing within 1 day, and the followup teams begin monitoring of 
the event immediately.  For example, the PLANET group generally produces 
template images the same day and is able to make rudimentary measurements 
of the event within a day after the receipt of the alert from the primary 
search teams.  Once these rudimentary measurements of the event are in place, 
one can determine a second crossing time without difficulties.  Therefore, 
the typical response time of the current followup observation teams is 2 
days if the weather is good (A.\ Gould 1999, private communication).  
According to the determined distribution $f(t_{\rm cc})$, then, the fraction 
of binary-lens events whose second caustic crossings can be monitored by 
the followup observations is $\sim 80\%$ although there exist  slight 
difference depending on the assumed mass functions.  Therefore, we find 
that for Galactic bulge events the detection bias toward long time-scale 
events argued by Honma (1999) will not be very strong.  However, we note 
that his argument about the detection bias against short time-scale events 
is based on the assumption that the current preparation time for intensive 
followup observations is greater than 7 days.  We are not certain why his 
adopted preparation time is that long.  However, if the same preparation 
time of 2 days is adopted, the fraction of LMC binary-lens events that can 
be intensively monitored by followup observations would be similar to that 
of Galactic bulge events.

However, the expected most frequent caustic-crossing interval of $\sim 1.5$ 
day is shorter than the current preparation time of the followup observations. 
Then, if the preparation time is shortened by adopting more aggressive 
observational strategy, one can detect and resolve caustic crossings for 
more binary-lens events.  With the help of automatized process, the MACHO 
group (Alcock et al.\ 1997) can finish data reduction 
within 6 hrs of image acquisition.  With vigorous efforts to minimize the 
preparation time, therefore, initiation of followup observations within 1 
day after the first caustic crossing will be possible.  We find that 
if the followup observations can be initiated within 1 day after the first 
caustic crossing, the detection rate can be improved into $\sim 90\%$.

\section*{Acknowledgments}
We would like to thank A.\ Gould for his helpful comments about the 
current strategy of microlensing followup observations.
This work was supported by the Korea Science \& Engineering Foundation
(grant 1999-2-113-001-5).

\end{document}